\documentclass[conference]{IEEEtran}

\usepackage{tikz}
\tikzset{>=latex}

\usepackage{color}
\usepackage{graphicx}
\usepackage{epstopdf}
\usepackage{amsmath}
\usepackage{amssymb}
\usepackage{hyperref}
\usepackage[english]{babel}
\usepackage{cite}
\usepackage{rotfloat}
\usepackage{mathtools}
\usepackage{amsmath}
\usepackage{makecell}
\usepackage{algorithm,algorithmic}
\usepackage{multirow}
\usepackage{subfigure}
\usepackage{booktabs}
\usepackage{colortbl}
\usepackage{multirow}
\usepackage{hhline}
\usepackage{stfloats}
\usepackage{multicol}
\usepackage[justification=centering,font=small,skip=0pt]{caption}

\newsavebox{\foobox}

\definecolor{kugray5}{RGB}{224,224,224}

\usepackage[normalem]{ulem}
\newcommand\rsout{\bgroup\markoverwith
	{\textcolor{red}{\rule[0.5ex]{2pt}{0.8pt}}}\ULon}



\makeatletter
\newcommand{\ALOOP}[1]{\ALC@it\algorithmicloop\ #1%
	\begin{ALC@loop}}
	\newcommand{\ENDALOOP}{\end{ALC@loop}\ALC@it\algorithmicendloop}

\makeatother

\usepackage{etoolbox}
\let\mybibitem\bibitem
\renewcommand{\bibitem}[1]{%
	\ifstrequal{#1}{nature}
	{\color{blue}\mybibitem{#1}}
	{\color{black}\mybibitem{#1}}%
}

\graphicspath{ {Figures/} }

\newtheorem{remark}{Remark}

\DeclareCaptionLabelSeparator{periodspace}{.\quad}

\addto\captionsenglish{}
\newcommand\nbthis{\addtocounter{equation}{1}\tag{\theequation}}

\newcommand{\abs}[1]{\left|#1\right|} 

\newcommand{\re}[1]{\mathfrak{R}{\left(#1\right)}}
\newcommand{\im}[1]{\mathfrak{I}{\left(#1\right)}}

\newcommand{\mean}[1]{\mathbb{E} \left\{#1\right\}}

\newcommand{\mH}{\textbf{\textit{H}}} 

\newcommand{\mA}{\textbf{\textit{A}}}

\newcommand{\mI}{\textbf{\textit{I}}}

\newcommand{\mB}{\textbf{\textit{B}}}

\newcommand{\mHr}{\textbf{\textit{H}}_r} 
\newcommand{\mHt}{\textbf{\textit{H}}_t} 
\newcommand{\mHd}{\textbf{\textit{H}}_d} 
\newcommand{\mIr}{\textbf{\textit{I}}_{N_r}}

\newcommand{\setC}{\mathbb{C}} 

\newcommand{\vx}{\textbf{\textit{x}}}
\newcommand{\vy}{\textbf{\textit{y}}}
\newcommand{\vr}{\textbf{\textit{r}}}

\newcommand{\vn}{\textbf{\textit{n}}}

\newcommand{\vh}{\textbf{\textit{h}}}

\newcommand{\vt}{\textbf{\textit{t}}}

\newcommand{\bPhi}{\boldsymbol{\Phi}}

\newcommand{\Pbs}{P_{\mathrm{BS}}}

\begin{document}
	
	\title{Machine Learning-based Reconfigurable Intelligent Surface-aided MIMO Systems}
	
	\author{\IEEEauthorblockN{Nhan~Thanh~Nguyen\IEEEauthorrefmark{1},
			Ly~V.~Nguyen\IEEEauthorrefmark{2}, Thien~Huynh-The\IEEEauthorrefmark{3},
			Duy~H.~N.~Nguyen\IEEEauthorrefmark{4},
			A.~Lee~Swindlehurst\IEEEauthorrefmark{5},\\ and
			Markku~Juntti\IEEEauthorrefmark{1}}
		\IEEEauthorblockA{\IEEEauthorrefmark{1}Centre for Wireless Communications, University of Oulu, P.O.Box 4500, FI-90014, Finland}
		\IEEEauthorblockA{\IEEEauthorrefmark{2}Computational Science Research Center, San Diego State University, CA, USA}
		\IEEEauthorblockA{\IEEEauthorrefmark{3}ICT Convergence Research Center, Kumoh National Institute of Technology, Gyeongsangbuk-do 39177, Korea}
		\IEEEauthorblockA{\IEEEauthorrefmark{4}Department of Electrical and Computer Engineering, San Diego State University, CA, USA}
		\IEEEauthorblockA{\IEEEauthorrefmark{5}Department of Electrical Engineering and Computer Science, University of California, Irvine, CA, USA.}
		\thanks{The research has been supported in part by Academy of Finland under 6Genesis Flagship (grant 318927) and EERA Project (grant 332362). 
			Corresponding author: N. T. Nguyen (email: nhan.nguyen@oulu.fi).}
		Email: \{nhan.nguyen, markku.juntti\}@oulu.fi, \{vnguyen6,  duy.nguyen\}@sdsu.edu, thienht@kumoh.ac.kr, swindle@uci.edu}

	\maketitle

	\begin{abstract}
		Reconfigurable intelligent surface (RIS) technology has recently emerged as a spectral- and cost-efficient approach for wireless communications systems. However, existing hand-engineered schemes for passive beamforming design and optimization of RIS, such as the alternating optimization (AO) approaches, require a high computational complexity, especially for multiple-input-multiple-output (MIMO) systems. To overcome this challenge, we propose a low-complexity unsupervised learning scheme, referred to as learning-phase-shift neural network (LPSNet), to efficiently find the solution to the spectral efficiency maximization problem in RIS-aided MIMO systems. In particular, the proposed LPSNet has an optimized input structure and requires a small number of layers and nodes to produce efficient phase shifts for the RIS. Simulation results for a $16 \times 2$ MIMO system assisted by an RIS with $40$ elements show that the LPSNet achieves $97.25\%$ of the SE provided by the AO counterpart with more than a $95\%$ reduction in complexity.
	\end{abstract}
	
	\begin{IEEEkeywords}
		Reconfigurable intelligent surface, RIS, IRS, MIMO, machine learning, passive beamforming.
	\end{IEEEkeywords}
	\IEEEpeerreviewmaketitle

	\section{Introduction}
	Reconfigurable intelligent surface (RIS) technology has recently been shown to be a promising solution for substantially enhancing the performance of wireless communications systems~\cite{di_renzo_smart_2019}. An RIS is often realized by a planar array comprising a large number of reconfigurable passive reflecting elements that can be configured to reflect incoming signals by a predetermined phase shift. Theoretical analyses have shown that an RIS of $N$ elements can achieve a total beamforming gain of $N^2$~\cite{wu2019intelligent} and an increase in the received signal power that increases quadratically with $N$~\cite{wang2020intelligent, Basar2019Reconfigurable}. 
	
	An important research direction for RIS-assisted communications systems is how to optimize the RIS reflecting coefficients to maximize the spectral efficiency (SE), e.g.,~\cite{gong2020towards,yang2020intelligent,yang2019irs, yuan2020intelligent, han2019large, di2020practical,zhang2020capacity,nguyen2021hybrid,nguyen2021spectral}. Yang \textit{et al.}~\cite{yang2020intelligent} jointly optimize the transmit power allocation and the passive  reflecting coefficients for an RIS-assisted single-input-single-output (SISO) system. By contrast, the work in \cite{yang2019irs, yuan2020intelligent, han2019large, di2020practical} considers multiple-input-single-output (MISO) systems assisted by RISs with continuous~\cite{yang2019irs, yuan2020intelligent} or coarsely quantized phase shifts~\cite{han2019large,di2020practical}. Efficient alternating optimization (AO) methods are developed in \cite{zhang2020capacity, nguyen2021hybrid, nguyen2021spectral} for the SE maximization problem of RIS-aided multiple-input-multiple-output (MIMO) systems.
	
	Machine learning (ML) has recently attracted much attention for wireless communication systems~\cite{pham2020intelligent, nguyen2020application, nguyen2020dnn}, and particularly for RIS, e.g.,~\cite{taha2019deep, Taha2020Deep,Feng2020Deep,Huang2020Reconfigurable,Gao2020Unsupervised,Ma2020Distributed}. The work in~\cite{taha2019deep, Taha2020Deep} considers SISO systems where the RIS contains some active elements. While a deep neural network (DNN) is used in~\cite{taha2019deep}, the work in~\cite{Taha2020Deep} exploits advances in deep reinforcement learning (DRL). Their solutions can approach the upper bound achieved with perfect channel state information (CSI). However, the introduction of active elements at the RIS causes additional hardware costs and power consumption. DRL is also exploited to optimize the RIS in assisting a system with a single-antenna user in~\cite{Feng2020Deep}, and for MIMO channels without a direct base station (BS) -- mobile station (MS) link~\cite{Huang2020Reconfigurable}. Gao \textit{et al.} \cite{Gao2020Unsupervised} propose a DNN-based passive beamforming (PBF) design, but only for a single-antenna user. Ma \textit{et al.}~\cite{Ma2020Distributed} exploit federated learning (FL) to enhance the PBF performance and user privacy, but again for a simplified system with a single-antenna BS and no direct link to the MS.
	
	Unlike the aforementioned studies on ML-based PBF, we consider a more general RIS-aided MIMO system in this paper. Furthermore, both the direct BS-MS and the reflected BS-RIS-MS links are taken into consideration, which imposes significantly more difficulty in the optimization of the ML models. In particular, the ML model needs to be optimized so that the signals through the different links add constructively at the MS. This implies another challenge, in that a large amount of information must be extracted by the ML model from the extremely high-dimension input data. Furthermore, unless an exhaustive search is performed, it is challenging to develop supervised learning-based ML models due to the unavailability of data labels. Fortunately, these challenges can be overcome by the findings in our paper. Specifically, by formulating the SE maximization problem and studying the AO solution, we discover an informative input structure that enables a DNN, even with a single hidden layer and an unsupervised learning strategy, to generate efficient phase shifts for PBF. The proposed DNN-based PBF scheme, referred to as a \textit{learning-phase-shift neural network (LPSNet)}, is numerically shown to perform very close to the AO method in~\cite{zhang2020capacity} with a substantial complexity reduction.
	
	\section{System Model and Problem Formulation}
	\label{sec_system_model}
	
	\subsection{System Model}
	
	
	We consider downlink transmission between a BS and MS that are equipped with $N_t$ and $N_r$ antennas, respectively. The communication is assisted by an RIS with $N$ passive reflecting elements. Let $\mHd \in \setC^{N_r \times N_t}$, $\mHt \in \setC^{N \times N_t}$, and $\mHr \in \setC^{N_r \times N}$ denote the BS-MS, BS-RIS, and RIS-MS channels, respectively, and let $\bPhi = \mathrm{diag} \{ \alpha_1, \alpha_2, \ldots, \alpha_N \}$ denote the diagonal reflecting matrix of the RIS. The RIS coefficients are assumed to have unit-modulus, i.e., $\alpha_n = e^{j \theta_n}, \forall n$, and, thus, it only introduces phase shifts $\theta_n \in [0, 2\pi), \forall n$ to the impinging signals. Let $\vx \in \setC^{N_t \times 1}$ be the transmitted signal vector. We focus on the design of the PBF and assume a uniform power allocation, i.e., $\mean{\vx \vx^H} = \Pbs \mI_{N_t}$, where $\Pbs$ is the transmit power at the BS. The MS receives the signals through the direct channel and via the reflection by the RIS. Therefore, the received signal vector at the MS can be expressed as
	\begin{align*}
		\vy &= \mHd \vx + \mHr \bPhi \mHt \vx + \vn = \mH \vx + \vn, \nbthis \label{eq_system_RIS}
	\end{align*}
	where $\mH \triangleq \mHd + \mHr \bPhi \mHt$ is the combined effective channel, and $\vn \sim \mathcal{CN} (0, \sigma^2 \mIr)$ is complex additive white Gaussian noise (AWGN) at the MS.
	
	\subsection{Problem Formulation}
	
	Based on \eqref{eq_system_RIS}, the SE of the RIS-aided MIMO system can be expressed as \cite{zhang2020capacity}
	\begin{align*}
		\mathrm{SE}\left( \{\alpha_n\} \right) =  \log_2 \mathrm{det} (\mIr + \rho \mH \mH^H), \nbthis \label{eq_SE}
	\end{align*}
	where $\{\alpha_n\} = \{\alpha_1, \alpha_2, \ldots, \alpha_N\}$ is the set of the phase shifts at the RIS that needs to be optimized, and $\rho = \frac{\Pbs}{\sigma^2}$. The PBF design maximizing the SE can be formulated as
	\begin{subequations}
		\label{prob_SE}
		\begin{align}
			(\mathrm{P0}) \quad \underset{\{\alpha_n\}}{\textrm{maximize}} \quad &  \mathrm{SE} \left( \{\alpha_n\} \right) \label{eq_objective} \\
			\textrm{subject to} \quad 
			& \abs{\alpha_n} = 1, \forall n. \label{cons_unit_modulus} 
		\end{align}
	\end{subequations}
	The objective function $\mathrm{SE} \left( \{\alpha_n\} \right)$, given in \eqref{eq_SE}, is nonconvex with respect to $\{\alpha_n\}$, and the feasible set for $(\mathrm{P0})$ is nonconvex due to the unit-modulus constraint \eqref{cons_unit_modulus}. Therefore, $(\mathrm{P0})$ is intractable and is difficult to find an optimal solution. Efficient solutions to $(\mathrm{P0})$ can be solved by the AO \cite{zhang2020capacity, nguyen2021spectral,nguyen2021hybrid} or projected gradient descent (PGD) \cite{perovic2020achievable} methods. However, these algorithms require an iterative update of $\{\alpha_n\}$. In particular, to obtain the phase shifts in each iteration, computationally expensive mathematical operations are performed, and as a result these algorithms have high complexity and latency. In the next section, we propose the low-complexity LPSNet to find an efficient solution to $(\mathrm{P0})$.
	
	\section{DNN-based PBF and Proposed LPSNet}
	
	\subsection{DNN-based PBF}
	
	Instead of employing a high-complexity algorithm (e.g., AO or PGD), a DNN can be modeled and trained to solve $(\mathrm{P0})$, i.e., to predict $\{\alpha_n\}$. While the phase shifts $\{\alpha_n\}$ are complex numbers and cannot be directly generated by a DNN, since $\abs{\alpha_n} = 1, \forall n$, the coefficients $\{\alpha_n\}$ are completely specified only by their real-valued phase shifts $\{\theta_n\}$. Hence, $(\mathrm{P0})$ is equivalent to
	\begin{align*}
		(\mathrm{P}) \quad \underset{\{\theta_n\}}{\textrm{maximize}} \quad &  \mathrm{SE} \left( \{\theta_n\} \right), \nbthis \label{prob_SE_1}
	\end{align*}
	 whose solution is found by a DNN in this work. 
	
	Let $\bar{\vh}$ be a vector containing the CSI parameters (i.e., the information in $\mHd$, $\mHt$, and $\mHr$). The phase shifts learned by a DNN can be mathematically modeled as
	\begin{align*}
		\{\hat{\theta}_n\} = f_{\mathrm{NN}} \left(  \bar{\vh} \right), \nbthis \label{eq_DNN}
	\end{align*}
	where $f_{\mathrm{NN}}(\cdot)$ represents the non-linear mapping from $\bar{\vh}$ to $\{\hat{\theta}_n\}$ performed by the DNN. The efficiency and structure of $f_{\mathrm{NN}}$ significantly depend on the input. Therefore, properly designing $\bar{\vh}$ is one of the first and most important tasks in modeling $f_{\mathrm{NN}}$. Let $\bar{\vh}_0$ be the input vector constructed from the  \textit{original CSI} (i.e., entries of $\mHd$, $\mHt$, and $\mHr$), i.e.,
	\begin{align*}
	    \bar{\vh}_0 \triangleq \mathrm{vec} \left( \mHd,\mHt,\mHr \right), \nbthis \label{eq_h0}
	\end{align*}
	where $\mathrm{vec}(\cdot)$ is a vectorization. Obviously, $\bar{\vh}_0$ contains the required CSI for a conventional hand-engineered algorithm to solve $(\mathrm{P0})$. However, $\bar{\vh}_0$ does not possess any structure or meaningful patterns that the DNN can exploit. Furthermore, if a very deep DNN is used, the resulting computational complexity may be as high as that of the conventional schemes, making it computationally inefficient. Our experimental results obtained by training DNNs with the input $\bar{\vh}_0$ show that the DNNs cannot escape from local optima even after extensive fine tuning. This motivates us to select more informative features as input of the proposed DNN.
	
	\subsection{Proposed Efficient Input Structure}
	
	To design a better-performing input structure for $f_{\mathrm{NN}}$, we first extract the role of each individual phase shift $\theta_n$ in $\mathrm{SE} \left( \{\theta_n\} \right)$. Let $\vr_n$ be the $n$th column of $\mHr$ and $\vt_n^H$ be the $n$th row of $\mHt$, i.e., $\mHr = [\vr_1, \ldots, \vr_N]$ and $\mHt = [\vt_1, \ldots, \vt_N]^H$. Furthermore, for ease of notation, define $\bar{\mH}_{(0)} \triangleq \mHd$ and $\theta_0 \triangleq 0$. Since $\bPhi$ is diagonal, one has
	\begin{align*}
		\mH = \mHd + \sum_{n=1}^N e^{j \theta_n} \vr_n \vt_n^H = \sum_{i=0}^N e^{j \theta_i} \bar{\mH}_{(i)}, \nbthis \label{eq_eff_channel_2}
	\end{align*}
	where $\bar{\mH}_{(n)} \triangleq \vr_n \vt_n^H$, $n = 1,\ldots,N$. Consequently, $\mathrm{SE} \left( \{\theta_n\} \right)$ can be recast in an explicit form
	\begin{align}
		\mathrm{SE}(\theta_n) = \log_2 \abs{\mA_n + e^{j \theta_n} \mB_n + e^{-j \theta_n} \mB_n^H}, \label{fm}
	\end{align}
	where
	\begin{align*}
		\mA_n &\triangleq \mIr + \rho \left(\sum_{i=0, i \neq n}^{N} e^{j \theta_i} \bar{\mH}_{(i)} \right) \left(\sum_{i=0, i \neq n}^{N} e^{j \theta_i} \bar{\mH}_{(i)} \right)^H \\
		&\hspace{5cm}  + \rho \bar{\mH}_{(n)} \bar{\mH}_{(n)}^H ,  \nbthis \label{def_A}\\
		\mB_n &\triangleq \rho \bar{\mH}_{(n)} \left( \sum_{i=0, i \neq n}^{N} e^{j \theta_i} \bar{\mH}_{(i)} \right)^H, \nbthis \label{def_B}
	\end{align*}
	$n=1,\ldots,N$. It is observed that neither $\mA_n$ or $\mB_n$ involves $\theta_n$. This means that if all variables $\{\theta_i\}_{i=0,i \neq n}^{N}$ are fixed, $\mA_n$ and $\mB_n$ are determined, and $\theta_n$ can be found by \cite{zhang2020capacity}
	\begin{align*}
		\theta_n = \mathrm{arg} \{ \lambda (\mA_n^{-1} \mB_n) \}, \nbthis \label{eq_sol_theta}
	\end{align*}
	where $\lambda (\mA_n^{-1} \mB_n)$ denotes the sole non-zero eigenvalue of $\mA_n^{-1} \mB_n$ \cite{zhang2020capacity}. This implies that the information necessary to obtain $\theta_n$ is contained in $\mA_n^{-1} \mB_n$, which can be computed if the values $\{\bar{\mH}_{(i)} \}$ are available, as observed in \eqref{def_A} and \eqref{def_B}. We summarize this important result in the following remark.
	\begin{remark}
	\label{rm_input}
		In a DNN predicting the phase shifts $\{ \theta_n \}$ of the RIS, the input vector should contain the coefficients of $\{\bar{\mH}_{(i)} \}$. More specifically, it can be constructed as
		\begin{align*}
			\bar{\vh} = \mathrm{vec} \left( \left\{ \re{ \bar{\mH}_{(i)} } \right\}, \left\{\im { \bar{\mH}_{(i)} } \right\} \right) \in \mathbb{R}^{2N_tN_r(N+1) \times 1}, \nbthis \label{eq_input_V}
		\end{align*}
		where $\re{\mA}$ and $\im{\mA}$ represent the real and imagine parts of the entries of $\mA$, respectively. There are $N+1$ matrices $\bar{\mH}_{(i)} \in \setC^{N_r \times N_t}$. Thus, $\bar{\vh}$ consists of $2N_tN_r(N+1)$ real elements.
	\end{remark}
	
	Some interesting observations can be made from \eqref{eq_eff_channel_2} and \eqref{eq_input_V}, as noted in the following remark.
	\begin{remark}
	     The efficient input structure of Remark~\ref{rm_input} is constructed from the pairwise products of the $N$ columns of $\mHr$ and the $N$ rows of $\mHt$. This is of interest because there are various ways to extract/combine the information from the entries of $\mHt$ and $\mHr$. Based on Remark \ref{rm_input}, it is reasonable to expect that the raw CSI $\bar{\vh}_0$ is not readily exploited by the DNN to learn our regression model successfully. Furthermore, $\mHd$, i.e., $\{\bar{\mH}_{(0)}\}$ in \eqref{eq_input_V}, contributes to $\bar{\vh}$ with its original entries, and no manipulation is required.
	\end{remark}
	
	 In addition, we note that although additional mathematical operations are required to obtain $\{\bar{\mH}_{(i)} \}$, they are just low-complexity multiplications of a column and a row vector and matrix additions. Therefore, generating the proposed input structure requires low computational complexity, but it has a significant impact on the learning ability and structure of the DNN employed for PBF, as will be shown in the next section.
	
	\subsection{Proposed LPSNet}
	\label{sec_LPSNet}
	
	Here we propose a learning-phase-shift neural network, referred to as LPSNet, for solving the PBF problem $(\mathrm{P})$ in~\eqref{prob_SE_1}. We stress that the informative input structure in \eqref{eq_input_V} plays a deciding role in the efficiency of the LPSNet. Specifically, $\bar{\vh}$ in \eqref{eq_input_V} already contains the meaningful information necessary to obtain $\{ \theta_n \}$. The network structure, training strategy, and computational complexity of LPSNet are presented next.

	\subsubsection{Network Structure}

	 LPSNet has a fully-connected DNN architecture, and like any other DNN model, it is optimized by fine-tuning the number of hidden layers and nodes, and the activation function. However, thanks to the handcrafted feature selection, the LPSNet can predict $\{ \theta_n \}$ efficiently with only a small number of layers. Through fine-tuning, we have found that only a single or a few hidden layers with $N$ nodes in each is sufficient to ensure satisfactory performance. As a result, the sizes of the input, hidden, and output layers are $2N_tN_r(N+1)$ (i.e., the size of $\bar{\vh}$), $N$, and $N$, respectively. Furthermore, we have found that Sigmoid activation functions at both the hidden and output layers can provide better training of the LPSNet than other activation functions. With such a shallow network, the complexity of LPSNet is very low. Moreover, we have found that to optimize the LPSNet, fine-tuning can be done by slightly increasing the number of hidden layers and nodes, e.g., for large MIMO systems. For example, to predict the phase shifts of an RIS equipped with $N=40$ elements assisting an $8 \times 2$ MIMO system, a single hidden layer with $N$ nodes can guarantee an efficient training, while for a $16 \times 2$ MIMO system, two hidden layers, each deployed with $N$ nodes, is sufficient (see Table~\ref{tab_LPSNet_config}). For other systems, it is recommended to begin fine-tuning with such a proposed simple network to avoid overfitting.
	
	\subsubsection{Training Strategy}
	 For offline training, we propose employing an unsupervised training strategy \cite{Gao2020Unsupervised}. We note that, on the other hand, if supervised training is used, labels ($\{\theta_n\}$) have to be obtained using a conventional high-complexity method such as AO \cite{zhang2020capacity} or PGD \cite{perovic2020achievable}. To avoid this high computational load, unsupervised training is adopted for training the LPSNet so that it can maximize the achievable SE without the labels $\{\theta_n\}$. To this end, we set the loss function to
	\begin{align*}
		\mathrm{Loss} (\{\hat{\theta}_n\}) = -\log_2 \mathrm{det} (\mIr + \rho_{\mathrm{train}} \mH_{\mathrm{train}} \mH_{\mathrm{train}}^H), \nbthis \label{eq_loss}
	\end{align*}
	where $\rho_{\mathrm{train}}$ is randomly picked from the range $[-\rho_0,\rho_0]$ dB, and $\mH_{\mathrm{train}} = \mH_{d,\mathrm{train}} + \mH_{r,\mathrm{train}} \hat{\bPhi} \mH_{t,\mathrm{train}}$. Here, $\mH_{d,\mathrm{train}}$, $\mH_{t,\mathrm{train}}$, and $\mH_{r,\mathrm{train}}$ are obtained by scaling $\mHd$, $\mHt$, and $\mHr$ by their corresponding large-scale fading coefficients so that their entries have zero-mean and unit-variance. This pre-processing is also applied to obtain the input data in $\bar{\vh}$. In other words, $\bar{\vh}$ is obtained from $\mH_{d,\mathrm{train}}$, $\mH_{t,\mathrm{train}}$, and $\mH_{r,\mathrm{train}}$ rather than $\mHd$, $\mHt$, and $\mHr$. Furthermore, $\hat{\bPhi} = \mathrm{diag} \{ e^{j \hat{\theta}_1}, \ldots, e^{j \hat{\theta}_N} \}$, where $	\{\hat{\theta}_n\}$ are the outputs of the LPSNet, as given in \eqref{eq_DNN}. During training, the weights and biases of the LPSNet are optimized such that $\mathrm{Loss}$ is minimized, or equivalently, SE is maximized.
	
	\subsubsection{Online LPSNet-based PBF}
	Once LPSNet has been trained, it is ready to be used online for PBF. At first, $\bar{\vh}$ is constructed from the CSI based on Remark \ref{rm_input}. Then, the LPSNet takes $\bar{\vh}$ as the input for output $\{ \hat{\theta}_n \}$, which are then transformed to the reflecting coefficients $\{ \hat{\alpha}_n \}$, with $\hat{\alpha}_n = \cos{\hat{\theta}} + j\sin{\hat{\theta}}, \forall n$. At this stage, the PBF matrix is obtained as $\boldsymbol{\Phi} = \mathrm{diag} \{ \hat{\alpha}_1 , \ldots, \hat{\alpha}_N\}$.
	
	\subsubsection{Computational Complexity Analysis}
	Let $L$ denote the number of hidden layers of LPSNet. LPSNet requires a complexity of $\mathcal{O}(N_rN_tN)$ for constructing the input $\bar{\vh}$, $\mathcal{O}(N_rN_tN^2)$ for the first hidden layer, and $\mathcal{O}(N^2)$ for subsequent hidden layers and the output layer. Therefore, the total complexity of the proposed LPSNet is $\mathcal{O}(\max(N_rN_tN^2,LN^2))$. When $L<N_rN_t$, the complexity of LPSNet is $\mathcal{O(}N_rN_tN^2)$. This should be the case since we found that LPSNet requires a small $L$ to produce efficient phase shift values; for example in Table~\ref{tab_LPSNet_config}, it is seen that only $1$ or $2$ hidden layers are required.
	
	For comparison, the computational complexity of the AO method in~\cite{zhang2020capacity} is $\mathcal{O}\big(N_rN_t(N + M)K + ((3N_r^3 + 2N_r^2N_t + N_t^2)N + N_rN_tM)I \big)$ where $M = \min(N_r,N_t)$, $K$ is the number of random sets $\{\alpha_n\}$ initializing the AO method, and $I$ is the number of iterations. Since $I$ is polynomial over $N_r$, $N_t$, and $N$, the total complexity of the AO method is at least quadratic in $N_r$, cubic in $N_t$, and quadratic in $N$. The complexity of LPSNet is only linear in $N_r$ and $N_t$, and quadratic in $N$. Therefore, LPSNet is less computationally demanding than the AO method. The computational complexity for each iteration of the PGM method in~\cite{perovic2020achievable} is
	$\mathcal{O}(2NN_tN_r + 2N_t^2 N_r + \frac{3}{2}N_tN_r^2 + N_r^3 + N_rN + N_tN + 3N + \frac{3}{2}N_t^3)$. Similarly, when the number of iterations is large, the PGM method is also more computationally expensive than LPSNet.
	
	\section{Simulation Results}
	\label{sec_sim_results}
	In this section, we numerically investigate the SE performance and computational complexity of the proposed LPSNet. We assume that the BS, MS, and RIS are deployed in a two-dimensional coordinate system at $(0,0)$, $(x_{\mathrm{MS}},y_{\mathrm{MS}})$, and $(x_{\mathrm{RIS}},0)$, respectively. The path loss for link distance $d$ is given by $\beta(d) = \beta_0 (d/1\mathrm{ m} )^{\epsilon}$, where $\beta_0$ is the path loss at the reference distance of $1$ m, and $\epsilon$ is the path loss exponent \cite{wu2019intelligent, zhang2020capacity}. Following \cite{zhang2020capacity, wu2019intelligent, nguyen2021hybrid}, we assume a Rician fading channel model for the small-scale fading of all the involved channels, including $\mHd$, $\mHt$, and $\mHr$. For more details on the the small-scale fading, please refer to \cite{zhang2020capacity, wu2019intelligent, nguyen2021hybrid}. We consider two systems, namely, $8 \times 2$ MIMO and $16 \times 2$ MIMO, both of which are aided by a RIS with $N=40$ reflecting elements \cite{zhang2020capacity, wu2019intelligent}. The bandwidth is set to $10$ MHz, and the noise power is $-170$ dBm/Hz \cite{zhang2020capacity}. Let $\epsilon_{(\cdot)}$ and $\kappa_{(\cdot)}$ respectively denote the path loss exponent and Rician factor of the corresponding channel $\mH_{(\cdot)}$. We set $\{\epsilon_d,\epsilon_t,\epsilon_r\} = \{3.5,2,2.8\}$ \cite{wu2019intelligent}, and due to the large distance and random scattering on the BS-MS channel, we set $\kappa_d = 0$ \cite{wu2019intelligent}. By contrast, to show that the proposed LPSNet can generalize to perform well for various physical deployments of the devices, in the simulations the RIS is randomly located on the $x-$axis, at a maximum distance of $100$m from the BS. Then, the MS is randomly placed $2$m away from the RIS. This setting implies that the MS is near the RIS, as widely assumed in the literature \cite{wu2019intelligent, zhang2020capacity, nguyen2021spectral}. Consequently, $\kappa_t$ and $\kappa_r$ are both randomly generated. This procedure applies to the generation of both the training and testing data sets.
	
    \begin{table}[t!]
      \begin{center}
        \caption{LPSNets employed for $8 \times 2$ and $16 \times 2$ MIMO systems.}
        \label{tab_LPSNet_config}
        \begin{tabular}{|c|c|c|c|}
        \hline
          MIMO system & \makecell{No. of\\hidden layers} & \makecell{No. of nodes \\in hidden layers} & Batch size \\
          \hline
          \hline
          $8 \times 2$ MIMO & $1$ & $N$ & $10$ samples \\
          \hline
          $16 \times 2$ MIMO & $2$ & $N \times N$ & $20$ samples \\
          \hline
        \end{tabular}
      \end{center}
      \vspace{-6mm}
    \end{table}
    
	LPSNet is implemented and trained using Python with the Tensorflow and Keras libraries and an NVIDIA Tesla V100 GPU processor. For the training phase, we employ the Adam optimizer (with a decaying learning rate of $0.99$ and a starting learning rate of $0.001$) and the loss function \eqref{eq_loss} with $\rho_0 = 30$ dB. We train and test the performance of the LPSNets, summarized in Table \ref{tab_LPSNet_config}, for two MIMO systems, namely $8 \times 2$ MIMO and $16 \times 2$ MIMO. For the former, LPSNet has only one hidden layer with $N$ nodes, whereas for the latter, two hidden layers, each with $N$ nodes, are employed. The batch sizes for training these LPSNets are $10$ and $20$ samples,  respectively, selected from data sets of $40,$$000$ samples.
	
	\subsection{The Significance of Input Structure}
	\begin{figure}[t]
	\setlength{\belowcaptionskip}{-15pt}
		\includegraphics[scale=0.65]{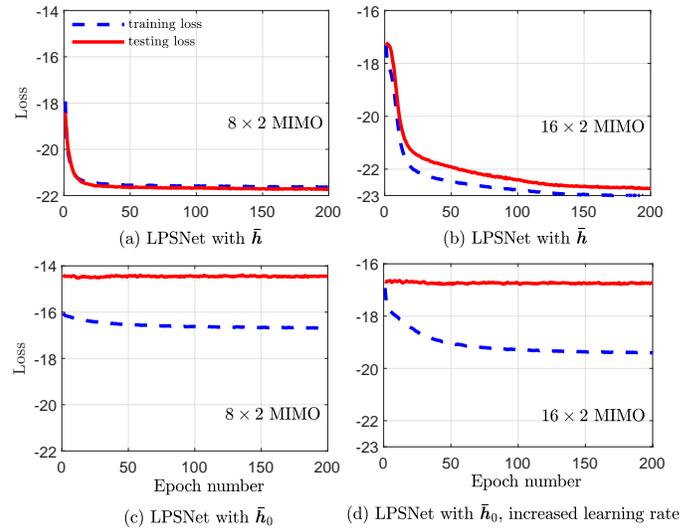}
		\caption{Loss of LPSNet when using the proposed input $\bar{\vh}$ and $\bar{\vh}_0$ for $8 \times 2$ MIMO and $16 \times 2$ MIMO systems.}
		\label{fig_loss}
	\end{figure}
	
	In Fig.\ \ref{fig_loss}, we justify the efficiency of the proposed input structure $\bar{\vh}$ in \eqref{eq_input_V} by comparing the resultant loss to that offered by $\bar{\vh}_0$ in \eqref{eq_h0}. It is observed for both the $8 \times 2$ MIMO (Fig.\ \ref{fig_loss}(a)) and $16 \times 2$ MIMO (Fig.\ \ref{fig_loss}(b)) systems that, as the epoch number increases, the loss value of the LPSNet with $\bar{\vh}$ quickly decreases until reaching convergence, with almost no or acceptable overfitting/underfitting. In contrast, when $\bar{\vh}_0$ is used, LPSNet cannot escape from the local optima, providing almost constant loss values (Figs.\ \ref{fig_loss}(c)). This is not overcome even when we increase the learning rate to $0.002$ and batch size to $100$ samples as in Fig.\ \ref{fig_loss}(d). The results in Fig.\ \ref{fig_loss} demonstrate that the input structure plays a key role in the learning ability of LPSNet, and the proposed informative input structure $\bar{\vh}$ enables LPSNet to be trained well for PBF even with simple architectures.
	
	\subsection{SE Performance and Complexity of the Proposed LPSNet}
	\begin{figure}[t]
	    \setlength{\belowcaptionskip}{-10pt}
		\includegraphics[scale=0.6]{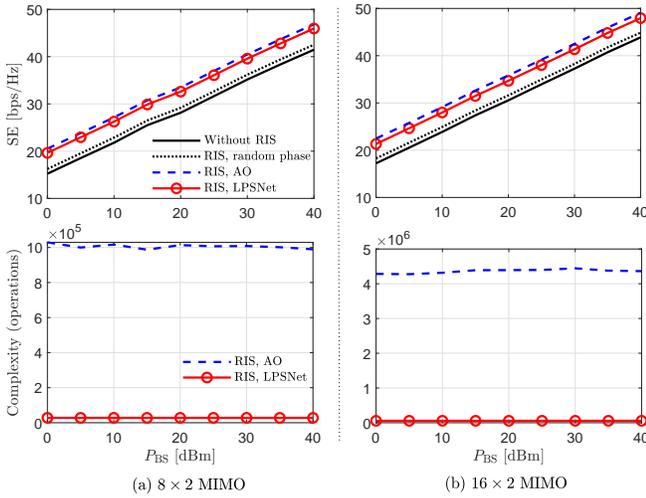}
		\caption{SE and complexity of the proposed LPSNet compared to other conventional schemes for (a) $8 \times 2$ and (b) $16 \times 2$ MIMO systems.}
		\label{fig_rate_comp}
	\end{figure}
	
	In Fig. \ref{fig_rate_comp}, the SE performance and computational complexity of the proposed LPSNet are shown for $8 \times 2$ and $16 \times 2$ MIMO systems. For comparison, we also consider the cases where $\{ \theta_n \}$ are either randomly generated, obtained based on the AO method \cite{zhang2020capacity}, or when RISs are not employed. For both considered MIMO systems we see that the proposed LPSNet performs far better than when using random phases for the RIS or without RIS. It performs similarly to the AO method but requires much lower complexity. For example, in the $16 \times 2$ MIMO system, LPSNet achieves $97.25\%$ of the SE provided by the AO approach with more than a $95\%$ reduction in complexity.
	
	\begin{figure}[t]
	    \setlength{\belowcaptionskip}{-20pt}
		\includegraphics[scale=0.6]{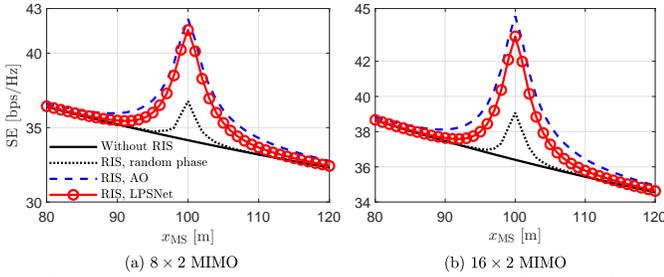}
		\caption{SE of the proposed LPSNet versus the positions of the MS for (a) $8 \times 2$ and (b) $16 \times 2$ MIMO systems with $P_{\mathrm{BS}} = 40$ dBm.}
		\label{fig_rate_d}
	\end{figure}
	
	In Fig.\ \ref{fig_rate_d}, we show the SE of the considered schemes when fixing $x_{\mathrm{RIS}} = 100$ m, $x_{\mathrm{MS}} \in [80,120]$ m, and $P_{\mathrm{BS}} = 40$ dBm. Clearly, the proposed LPSNet attains good performance over the entire considered range of $x_{\mathrm{MS}}$, which demonstrates its ability to perform well regardless of the physical distances between devices. This is because the training data (i.e., the channel coefficients) have been pre-processed to remove the effect of the large-scale fading. Furthermore, LPSNet provides the highest SE at $x_{\mathrm{MS}} = x_{\mathrm{RIS}} = 100$ m, i.e., when the RIS and MS are closest to each other, in agreement with the findings of \cite{wu2019intelligent, bjornson_intelligent_2019}.

	\section{Conclusion}
	\label{sec_concusion}
	In this paper, we have considered the application of DNN to passive beamforming in RIS-assisted MIMO systems. Specifically, we formulated the SE maximization problem, and showed how it can be solved efficiently by the proposed unsupervised learning-based LPSNet. With the optimized input structure, LPSNet requires only a small number of layers and nodes to output reliable phase shifts for PBF. It achieves almost the same performance as the AO method with considerably less computational complexity. Furthermore, the proposed unsupervised learning scheme does not require any computational load for data labeling.
	

	\bibliographystyle{IEEEtran}
	\bibliography{IEEEabrv,Ref}

\begin{thebibliography}{10}
\providecommand{\url}[1]{#1}
\csname url@samestyle\endcsname
\providecommand{\newblock}{\relax}
\providecommand{\bibinfo}[2]{#2}
\providecommand{\BIBentrySTDinterwordspacing}{\spaceskip=0pt\relax}
\providecommand{\BIBentryALTinterwordstretchfactor}{4}
\providecommand{\BIBentryALTinterwordspacing}{\spaceskip=\fontdimen2\font plus
\BIBentryALTinterwordstretchfactor\fontdimen3\font minus
  \fontdimen4\font\relax}
\providecommand{\BIBforeignlanguage}[2]{{%
\expandafter\ifx\csname l@#1\endcsname\relax
\typeout{** WARNING: IEEEtran.bst: No hyphenation pattern has been}%
\typeout{** loaded for the language `#1'. Using the pattern for}%
\typeout{** the default language instead.}%
\else
\language=\csname l@#1\endcsname
\fi
#2}}
\providecommand{\BIBdecl}{\relax}
\BIBdecl

\bibitem{di_renzo_smart_2019}
M.~D. Renzo, M.~Debbah, D.~T.~P. Huy, A.~Zappone, M.~Alouini, C.~Yuen,
  V.~Sciancalepore, G.~C. Alexandropoulos, J.~Hoydis, H.~Gacanin, J.~de~Rosny,
  A.~Bounceur, G.~Lerosey, and M.~Fink, ``Smart radio environments empowered by
  reconfigurable {AI} meta-surfaces: an idea whose time has come,''
  \emph{{EURASIP} J. Wireless Comm. Network.}, vol. 2019, p. 129, 2019.

\bibitem{wu2019intelligent}
Q.~Wu and R.~Zhang, ``Intelligent reflecting surface enhanced wireless network
  via joint active and passive beamforming,'' \emph{{IEEE} Trans. Wireless
  Commun.}, vol.~18, no.~11, pp. 5394--5409, 2019.

\bibitem{wang2020intelligent}
P.~Wang, J.~Fang, X.~Yuan, Z.~Chen, and H.~Li, ``{Intelligent reflecting
  surface-assisted millimeter wave communications: Joint active and passive
  precoding design},'' \emph{IEEE Trans. Veh. Tech.}, 2020.

\bibitem{Basar2019Reconfigurable}
E.~{Basar}, M.~{Di Renzo}, J.~{De Rosny}, M.~{Debbah}, M.~{Alouini}, and
  R.~{Zhang}, ``Wireless communications through reconfigurable intelligent
  surfaces,'' \emph{IEEE Access}, vol.~7, pp. 116\,753--116\,773, Sept. 2019.

\bibitem{gong2020towards}
S.~Gong, X.~Lu, D.~T. Hoang, D.~Niyato, L.~Shu, D.~I. Kim, and Y.-C. Liang,
  ``Towards smart wireless communications via intelligent reflecting surfaces:
  A contemporary survey,'' \emph{IEEE Commun. Surveys Tuts.}, 2020.

\bibitem{yang2020intelligent}
Y.~Yang, B.~Zheng, S.~Zhang, and R.~Zhang, ``{Intelligent reflecting surface
  meets OFDM: Protocol design and rate maximization},'' \emph{IEEE Trans.
  Commun.}, 2020.

\bibitem{yang2019irs}
Y.~Yang, S.~Zhang, and R.~Zhang, ``{IRS-enhanced OFDM: Power allocation and
  passive array optimization},'' in \emph{IEEE Global Commun. Conf.
  (GLOBECOM)}, 2019, pp. 1--6.

\bibitem{yuan2020intelligent}
J.~Yuan, Y.-C. Liang, J.~Joung, G.~Feng, and E.~G. Larsson, ``{Intelligent
  reflecting surface-assisted cognitive radio system},'' \emph{IEEE Trans.
  Commun.}, 2020.

\bibitem{han2019large}
Y.~Han, W.~Tang, S.~Jin, C.-K. Wen, and X.~Ma, ``{Large intelligent
  surface-assisted wireless communication exploiting statistical CSI},''
  \emph{IEEE Trans. Veh. Tech.}, vol.~68, no.~8, pp. 8238--8242, 2019.

\bibitem{di2020practical}
B.~Di, H.~Zhang, L.~Li, L.~Song, Y.~Li, and Z.~Han, ``{Practical Hybrid
  Beamforming With Finite-Resolution Phase Shifters for Reconfigurable
  Intelligent Surface Based Multi-User Communications},'' \emph{IEEE Trans.
  Veh. Tech.}, vol.~69, no.~4, pp. 4565--4570, 2020.

\bibitem{zhang2020capacity}
S.~Zhang and R.~Zhang, ``Capacity characterization for intelligent reflecting
  surface aided mimo communication,'' \emph{{IEEE} J. Sel. Areas Commun.},
  vol.~38, no.~8, pp. 1823--1838, 2020.

\bibitem{nguyen2021hybrid}
\BIBentryALTinterwordspacing
N.~T. Nguyen, Q.-D. Vu, K.~Lee, and M.~Juntti, ``{Hybrid relay-reflecting
  intelligent surface-assisted wireless communication},'' \emph{arXiv}, 2021.
  [Online]. Available: \url{https://arxiv.org/abs/2103.03900}
\BIBentrySTDinterwordspacing

\bibitem{nguyen2021spectral}
------, ``{Spectral efficiency optimization for hybrid relay-reflecting
  intelligent surface},'' \emph{to be published in IEEE Int. Conf. Commun.
  Workshops (ICCW)}, 2021.

\bibitem{pham2020intelligent}
Q.-V. Pham, N.~T. Nguyen, T.~Huynh-The, L.~B. Le, K.~Lee, and W.-J. Hwang,
  ``{Intelligent Radio Signal Processing: A Contemporary Survey},'' \emph{arXiv
  preprint arXiv:2008.08264}, 2020.

\bibitem{nguyen2020application}
N.~T. Nguyen, K.~Lee, and H.~Dai, ``{Application of Deep Learning to Sphere
  Decoding for Large MIMO Systems},'' \emph{arXiv preprint arXiv:2010.13481},
  2020.

\bibitem{nguyen2020dnn}
L.~V. Nguyen, D.~H. Nguyen, and A.~L. Swindlehurst, ``{DNN-based Detectors for
  Massive MIMO Systems with Low-Resolution ADCs},'' \emph{arXiv preprint
  arXiv:2011.03325}, 2020.

\bibitem{taha2019deep}
A.~{Taha}, M.~{Alrabeiah}, and A.~{Alkhateeb}, ``Deep learning for large
  intelligent surfaces in millimeter wave and massive {MIMO} systems,'' in
  \emph{Proc. IEEE Global Commun. Conf.}, Waikoloa, HI, USA, Dec. 2019.

\bibitem{Taha2020Deep}
A.~{Taha}, Y.~{Zhang}, F.~B. {Mismar}, and A.~{Alkhateeb}, ``Deep reinforcement
  learning for intelligent reflecting surfaces: {T}owards standalone
  operation,'' in \emph{Proc. IEEE Int. Workshop on Signal Process. Advances in
  Wireless Commun.}, Atlanta, GA, USA, May 2020.

\bibitem{Feng2020Deep}
K.~{Feng}, Q.~{Wang}, X.~{Li}, and C.~{Wen}, ``Deep reinforcement learning
  based intelligent reflecting surface optimization for {MISO} communication
  systems,'' \emph{IEEE Wireless Commun. Letters}, vol.~9, no.~5, pp. 745--749,
  May 2020.

\bibitem{Huang2020Reconfigurable}
C.~{Huang}, R.~{Mo}, and C.~{Yuen}, ``Reconfigurable intelligent surface
  assisted multiuser {MISO} systems exploiting deep reinforcement learning,''
  \emph{IEEE J. Select. Areas in Commun.}, vol.~38, no.~8, pp. 1839--1850, Aug.
  2020.

\bibitem{Gao2020Unsupervised}
J.~{Gao}, C.~{Zhong}, X.~{Chen}, H.~{Lin}, and Z.~{Zhang}, ``Unsupervised
  learning for passive beamforming,'' \emph{IEEE Commun. Letters}, vol.~24,
  no.~5, pp. 1052--1056, May 2020.

\bibitem{Ma2020Distributed}
D.~{Ma}, L.~{Li}, H.~{Ren}, D.~{Wang}, X.~{Li}, and Z.~{Han}, ``Distributed
  rate optimization for intelligent reflecting surface with federated
  learning,'' in \emph{Proc. IEEE Int. Conf. Commun. Workshops (ICCW)}, Dublin,
  Ireland, June 2020.

\bibitem{perovic2020achievable}
N.~S. Perovi{\'c}, L.-N. Tran, M.~Di~Renzo, and M.~F. Flanagan, ``{Achievable
  Rate Optimization for MIMO Systems with Reconfigurable Intelligent
  Surfaces},'' \emph{arXiv preprint arXiv:2008.09563}, 2020.

\bibitem{bjornson_intelligent_2019}
E.~Bj\"{o}rnson, O.~\"{O}zdogan, and E.~G. Larsson, ``Intelligent reflecting
  surface vs. decode-and-forward: How large surfaces are needed to beat
  relaying?'' \emph{IEEE Wireless Commun. Lett.}, pp. 1--1, 2019.

\end{thebibliography}

\end{document}